\documentclass{article}
\usepackage{arxiv}
\usepackage{lineno,hyperref}
\usepackage[figuresright]{rotating}
\usepackage{graphicx}
\usepackage{pdfpages}
\usepackage{float}
\usepackage{array, tabularx}
\usepackage{adjustbox}
\renewcommand*{\arraystretch}{1.5}
\usepackage{multirow}
\usepackage[lofdepth]{subfig}
\usepackage{algorithm}
\usepackage[noend]{algpseudocode}
\usepackage{comment}
\usepackage{setspace}
\usepackage{mathtools}
\usepackage{amsmath}
\modulolinenumbers[5]

\title{An Object Aware Hybrid U-Net for Breast Tumour Annotation}
\author{
 Suvidha Tripathi \\
  Department of Information Technology\\
  Indian Institute of Information Technology Allahabad\\
  Jhalwa, Deoghat, Prayagraj, Uttar Pradesh 211015 \\
  \texttt{suvitri24@gmail.com} \\
   \And
 Satish Kumar Singh \\
  Department of Information Technology\\
  Indian Institute of Information Technology Allahabad\\
  Jhalwa, Deoghat, Prayagraj, Uttar Pradesh 211015 \\
  \texttt{sk.singh@iiita.ac.in} \\
}
\begin{document}
\maketitle

\begin{abstract}
In the clinical settings, during digital examination of histopathological slides, the pathologist annotate the slides by marking the rough boundary around the suspected tumour region. The marking or annotation is generally represented as a polygonal boundary that covers the extent of the tumour in the slide. These polygonal markings are difficult to imitate through CAD techniques since the tumour regions are heterogeneous and hence segmenting them would require exhaustive pixel wise ground truth annotation. Therefore, for CAD analysis, the ground truths are generally  annotated by pathologist explicitly for research purposes. However, this kind of annotation which is generally required for semantic or instance segmentation is time consuming and tedious. In this proposed work, therefore, we have tried to imitate pathologist like annotation by segmenting tumour extents by polygonal boundaries. For polygon like annotation or segmentation, we have used Active Contours whose vertices or snake points move towards the boundary of the object of interest to find the region of minimum energy. To penalize the Active Contour we used modified U-Net architecture for learning penalization values. The proposed hybrid deep learning model fuses the modern deep learning segmentation algorithm with traditional Active Contours segmentation technique. The model is tested against both state-of-the-art semantic segmentation and hybrid models for performance evaluation against contemporary work. The results obtained show that the pathologist like annotation could be achieved by developing such hybrid models that integrate the domain knowledge through classical segmentation methods like Active Contours and global knowledge through semantic segmentation deep learning models.
\end{abstract}


\section{Introduction}
\label{C7S1}
Accurate segmentation of biological structures and micro-structures visualized in a digitized biopsy images could assist pathologists in measuring the disease extent. It could also help biomedical researchers around the world by automatically annotating the huge amount of medical images. In recent times, with arrival of deep learning techniques, the methods for image analysis have advanced rapidly, more so with easy availability of vast amounts of data such as ImageNet, Cifar100, COCO, etc. for various applications. The architectures trained on such huge amounts of data set a benchmark for similar applications. However, no state-of-the-art deep learning model except U-Net \cite{ronneberger2015u} has been proposed for biomedical image analysis because of less data volume in medical domain, high variability among datasets, and various modalities to address. The huge variations and less annotated data prevents the generalization of deep learning models. In such a scenario, most of the models proposed in this domain use pre-trained deep architectures or modify existing ones to suit their application. For segmentation problem, U-Net has been used extensively as base architecture \cite{li2017multi,qu2019joint,alom2018nuclei,sirinukunwattana2017gland,graham2019mild,zeng2019ric,
bentaieb2016topology,cui2018deep,oda2018besnet}. U-Net is basically a semantic segmentation architecture. Semantic segmentation classifies each pixel in a target region as a class or non-class pixel.  It is a supervised learning model whose performance depends on how well the ground truth is annotated. Ground truth annotation however in case of multi-structural histopathological images is a tedious task. This is where U-Net gives the advantage as it is known to produce better results even with limited datasets. The U-Net architecture follows the basic encoder-decoder structure with contracting path on the left and expanding path on the right. The unique difference between encoder-decoder architecture and U-Net makes the U-Net network more robust with limited dataset. Besides contracting-expanding modules, U-Net preserves the feature lost during contracting path by concatenating them with the expanding feature map in the expanding path of the model. The additional features preserves the quality of the segmentation. 
\par The dataset used for our work comprise Whole Slide Images with three annotated tumour classes- benign, invasive, and \textit{In situ}. The classes are very roughly annotated by experts to indicate the presence of the disease. The problem aggravates in case of invasive carcinoma where it is difficult to draw a boundary to contain the class. High grade invasive carcinoma could spreads across the WSI and does not have a epithelial layer boundary to contain the malignant cells. In such cases, accurate segmentation becomes a challenge. 
Hence, we aim to develop a rough segmentation framework to indicate the presence of the tumour in the region. Rough segmentation of complex tumour regions is an open research problem. The pathologists themselves do not extensively annotate cells and nuclei to detect a cancerous region, instead they annotated a rough boundary around the suspected region to mark the presence of the disease. These rough boundaries often also contain the cluster of small tumours that are spread across the slide and cannot be bounded separately for each cluster. Hence, such clusters are generally annotated as single region bounded by a rough annotation. The region of interest thus also, often contain non-tumourous portions that are found in-between or around the clusters.  In such cases, semantic or instance segmentation models often fails to precisely separate objects from background. Hence, we need to develop models that could know about the possible object location before segmentation.
\par Therefore, auto-initialized active contours which are initialized using a constrained criteria could be useful for such tasks. Active contours \cite{chan2001active} are known to predict high-level object shapes by finding the possible boundary of the object depending on both the image features and priors such as length and curvature of the contour, and other forces that drive the contour towards the edge of the object. These local priors are selected depending upon the application. Active contours find a minimum energy fit for incrementing contour vertices to the object edges. The energy is defined in such a way that the contours are attracted towards the minima or where the boundaries of the object lies. Active contours have an advantage because they are topology aware and could also work on high-level image features acquired at low resolution. In our proposed method, due to space constraints, high resolution WSI regions were downsampled to low-resolution small and even dimension images for input into modified U-Net architecture.   
\par In the proposed work, we tried to amalgamate U-net and active contours to develop a object aware segmentation network for segmenting breast tumour images belonging to three different image categories. We modified U-Net network using deep neatwork tools such as ResNet \cite{he2016deep} and DenseNet \cite{huang2017densely} blocks to enhance the efficiency of the segmentation. The U-net base acts as a learning framework for active-contour priors that are responsible for length and curvature of the contours. The priors are learned while the contour moves with each epoch. The active learning model generates polygons close to ground truth instance. The network is inspired from the original work \cite{marcos2018learning} that uses custom CNN and Active Contour Model (ACM). They used structured prediction for optimizing ACM parameters and SSVM (structured SVM) loss for finding optimal parameters. We tested our method using Intersection over Union (IoU) metric with the original article along with other benchmark methods on breast tumour dataset. Our method outperformed all the methods used for similar task. 
\par We have presented a preliminary analysis on the proposed method and more detailed analysis is our future work. 
The main contributions of the proposed work  are: \begin{itemize}
\item The novelty of the work lies in its application in histopatholgical images. We tried to achieve the pathologist like annotation through this method. 
\item We achieved the active contour inference with deep learning training on the roughly annotated tumour dataset in an end-to-end network
\item The successful implementation of the method and the comparison with contemporary state-of-the-art methods highlights the importance of integrating traditional methods with deep learning methods for better results.
\end{itemize} 
\section{Related Work}
\label{C7S2}

In digital pathology, tissue-wise labelled data is limited because it is very time consuming and requires expert pathologists. Hence, segmentation outputs are noisy and affects classification performance if the framework is end-to-end. Inconsistency in data acquisition methods also make limited amount of data even more useless. Due to these limitations, not much work has been done recently for segmenting tissue level regions. The authors in \cite{mehta2018net} have done a similar work in which they have taken four classes of breast biopsy tumours, namely, benign, atypia, DCIS and invasive. Their work divides WSIs into instances for feeding them into their network for joint segmentation and classification task. The output of their model produces an instance level segmentation mask and instance level probability map. The combined discriminative segmentation mask from the two outputs is then used to extract frequency and  co-occurrence features which were then fed into MLP for final cancer diagnosis. The strength of their work is that they have used general UNeT \cite{ronneberger2015u} architecture for their specific task using simple modifications like adding instance level probability map to enhance the features of segmentation mask that helped in improving classification accuracy of the final diagnosis. Their dataset was heavily annotated with tissue-level annotation done by 87 pathologists along with extensive substructures annotation by a pathology fellow. Exhaustive annotation is one of their most important strengths that aided in producing less noisy segmentation masks. This also helped them to create an end-to-end learning framework for both of their tasks. However, this is also the main drawback, that without the heavy annotations their method would not work. The similar BACH dataset whcih we have used in the proposed work , when tested on their algorithm failed to produce comparable results. Needless to say that other medical data segmentation algorithms like UNeT \cite{ronneberger2015u}, SegNet \cite{badrinarayanan2017segnet}, FCN \cite{long2015fully} also failed to perform well on our dataset due to the same limitation. These pixel classification based segmentation methods have a fundamental limitation when the target object comprise of many heterogeneous components. For example, a benign tumour at low resolution not only consists of nuclei, but structures like papillary, solid, hemorrhagic, and sclerotic growth patterns are also visible. They are surrounded by an well formed one-two layers of epithelium cells and fibrous sheath of connective tissue. At higher magnification, one can see solid areas composing stromal cells, round nuclie with fine chromatin and rare nucleoli \cite{ackerman1971pathology}. Such varied structures in one tumour cannot be individually annotated for semantic segmentation and hence treating them as one structure as a whole poses great confusion for such algorithms and therefore, fail to produce good results. The problem with instance segmentation algorithms like MaskRCNN \cite{he2017mask} is that it requires complete object to be present in the image for segmentation and classification. Region Proposal network of MaskRCNN compares the object characteristics as a whole, with certain threshold,  with the learned instances to propose a probable bounding box. Whereas in case of medical histology images, object characteristics vary widely within classes, and it gets very difficult to learn all types of object priors for smooth detection and classification. Other recent detection and classification methods like FastRCNN \cite{girshick2015fast} and \cite{ren2015faster} also works on the same theory of instance level detection and classification. Other works that our similar to ours are the segmentation based method in \cite{mehta2018learning} and saliency map-based method in \cite{geccer2016detection}.  Mehta et al. \cite{mehta2018learning} developed a CNN-based method for segmenting breast biopsy images that produces a tissue-level segmentation mask for each WSI. The histogram features they extracted from the segmentation masks were used for diagnostic classification. Geccer et al. \cite{geccer2016detection} proposed a saliency-based method for diagnosing cancer in breast biopsy images that identified relevant regions in breast biopsy WSIs to be used for diagnostic classification. 
\par Due to the rough annotation, our work focuses not only on segmenting the tumour mask from the background but also tried to draw a contour around the mask for better object boundary visualization. The similar literatures in the past have termed such tasks as contour-aware segmentations \cite{kainz2015semantic,chen2016dcan}. Classically, active contours have been extensively used in histopathology images for segmenting nuclei and cells \cite{xu2010weighted,xu2011high,ali2012integrated}. But, using ACM with CNN and for larger tissue regions like glands remains under-explored. Recently, Xu et al. \cite{xu2019convolutional} segmented nuclei from breast biopsy histopathological images that uses CNN for nuclei detection and the detected nuclei act as a initialization for active contour based ellipse fitting over the  detected nuclei.  \cite{khvostikov2019trainable} trained the CNN model for learning active contour priors for gland segmentation. They also proposed a collision resolution algorithm as a post processing step to separate overlapping gland objects. The dataset used by them was carefully annotated with crisp gland boundaries. However, in our case, since the annotations are rough and there is no crisp boundaries for tumours, the task of segmentation becomes even more challenging. Therefore, in the proposed work, we have preliminary studies the performance of our model on the dataset and compared it with recent segmentation benchmarks. 

\section{Methodology}
\label{C7S3}
\begin{sidewaysfigure}[htbp]
\centering
\includegraphics[width=\textwidth]{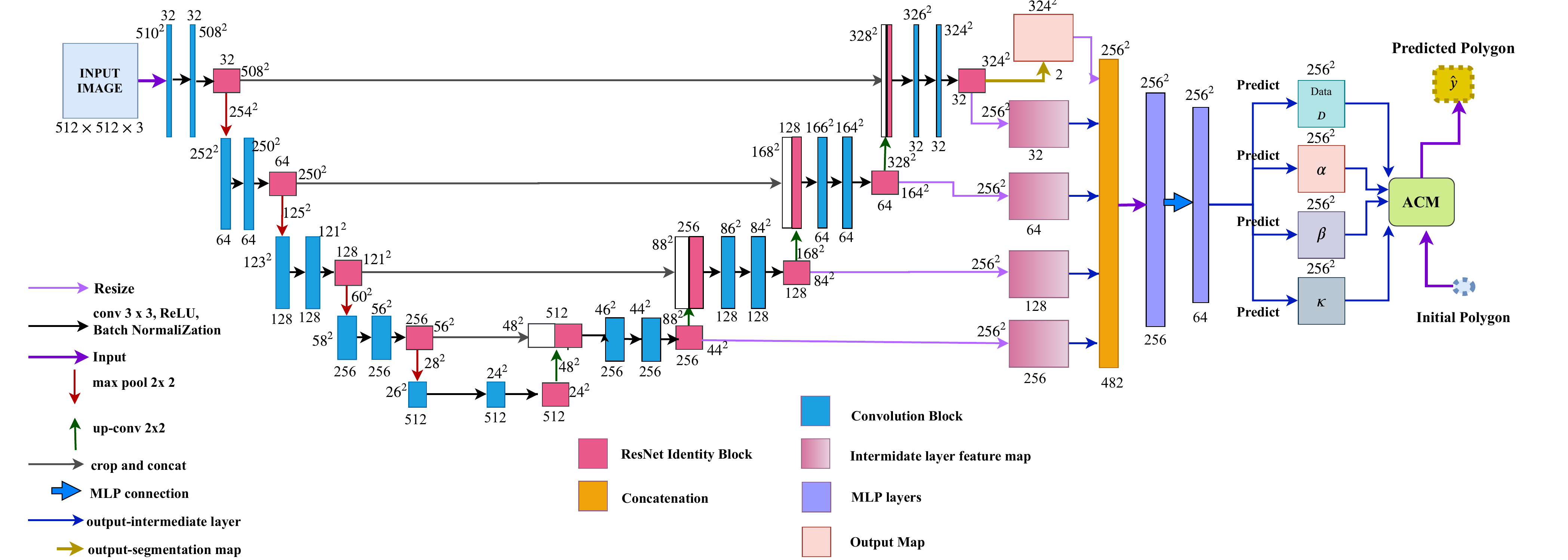}
\caption{Block Diagram of proposed hybrid segmentation model}
\label{C7F1}
\end{sidewaysfigure}
\subsection{Overview}
\label{C7S3S1}
The proposed model segments the roughly annotated breast tumour masks using hybrid U-Net Active Contour model. The backbone U-net model is modified by adding ResNet blocks. The detailed modified U-Net model is illustrated in \autoref{C7F1}. The output map of each up-sampling layer is concatenated to produce a concatenated feature map. This feature map is then further processed with convolution layers to produce four active contour priors. Each of the feature prior is then used to calculate active contour energy terms. The active contour energy function equation (refer \autoref{eq1chap7}) has four local priors, i.e. values that weigh the contour energy terms on a per-pixel basis. Hence, the priors as calculated as feature maps which are dynamically learned during back-propagation in an end-to-end model. The block diagram of the proposed model is shown in \autoref{C7F1}. We have modified the methodology used in DSAC \cite{marcos2018learning}. The difference is that their method uses CNN to learn Active Contour Model external and internal energy terms \cite{chan2001active} while we have used modified U-net architecture to get the deep learning inference (\autoref{C7S3S2}). Furthermore, the strategy to use automatic multiple initializations has been expanded so that the initial contour can find the object even if it is not centred (refer to \autoref{C7S3S4}). The modified training algorithm is expressed in Algorithm \ref{alg0chap7_1}.
\setlength{\textfloatsep}{1pt}
\begin{algorithm}[htbp]
\caption{proposed model training algorithm. The deep learning backbone forwards the feature maps learned to ACM inference at every iteration. Then ACM inference is made for each initialized polygon. The five polygons yield five polygons. A structured loss IoU is calculated for all the five polygons. After the training is complete, the maximum IoU is calculated to yield the final polygon}
\begin{algorithmic}
\State $\mathbf{Data:} \gets X,Y:$ \textit{image/polygon pairs in the training set} \par
\State $\mathbf{y^0} \gets$ \textit{corresponding polygon initializations} \par
\For{$x_i, y_i \in X, Y$}\par
    \State UNet-ResNet inference: $D$, $\alpha$, $\beta$, $\kappa \gets CNN_w(x_i)$ \par
    \State ACM inference: \par
    \For {$j=1:n$} \par 
      \State ${\hat{y}}^{j}_{i} \gets ACM(D$, $\alpha$, $\beta$, $\kappa$, $y^{j}_{i})$ \par
      \State $\frac{\partial{L^j}}{\partial{D}}$, $\frac{\partial{L^j}}{\partial{\alpha}}$,                                            $\frac{\partial{L^j}}{\partial{\beta}}$, $\frac{\partial{L^j}}{\partial{\kappa}} 
             \gets {\hat{y}}^{j}_{i}$,$y^{j}_{i}$ \par
      \State Compute $\frac{\partial{L^j}}{\partial{\omega}}$ (combined loss) using backpropagation \par 
      \State Update UNet-ResNet: $\omega \gets \omega - \eta(\frac{\partial{L^j}}{\partial{\omega}})$ \par 
      \State Calculate $IoU^j$
    \EndFor
    \State Determine max IoU and corresponding index
\EndFor
\end{algorithmic}
\label{alg0chap7_1}
\end{algorithm}

\subsection{Modified U-Net Architecture}
\label{C7S3S2}
Original U-net architecture \cite{ronneberger2015u} has an encoding and a decoding branch comprising a stack of convolution and deconvolution blocks, respectively. The encoder branch learns input representations while downsampling the input image, whereas the decoder branch recovers the spatial resolution lost during downsampling. The spatial information lost due to the downsampling of the input is added back at the up-sampling layer using the skip connections. These skip connections are made between corresponding layers of encoder and decoder branch. We have added resnet identity blocks after the convolution blocks in each layer on both encoder and decoder branches. The resnet blocks further help recover spatial information loss in the whole model. 
In the decoder branch, the output feature maps from each layer are then resized to the output size ($256 \times 256$) and concatenated to produce the final output feature map. This feature map is then further passed through two-layer MLP with 256 and 64 hidden units to predict four local information priors or weight maps: Data $D(x)$, $\alpha(x)$, $\beta(x)$, and $\kappa(x)$, where $x$ is the input image; $x \in X$. The active contour inference and the prediction of local priors are followed as in the literature \cite{ronneberger2015u}. 
\subsection{Active Contour}
\label{C7S3S3}
An active contour \cite{chan2001active} is a line or a continuous set of points that move over the image to find the point of minima. In other words, each point in a contour moves around the image so that the energy function is minimized. 
An active contour can be represented as a polygon $y=(u,v)$ with $L$ nodes. Let each node $s$ is represented by $y_s=(u_s, v_s)$ with $s\in 1,\ldots,L$. The polygon $y$ is then deformed such that the following energy function is minimized. 
\begin{equation}
\label{eq1chap7}
\begin{aligned}
E(y) &=\sum^{L}_{s=1}[D(y_s)+\alpha(y_s)\mid\frac{\Delta_s y_s}{\Delta s}\mid^2 + \\
       & \beta(y_s)\mid\frac{\Delta^2_s y_s}{\Delta s^2}\mid^2] + \sum_{u,v \in \Omega{(y)}} \kappa{(u,v)}
\end{aligned}
\end{equation}  
$D(y_s)$ is the external energy term indexed by the position $y_s=(u_s, v_s)$ and means the value in function $D(x)$, where $x$ is the input image,  where $D(x) \in \Re^(U\times V)$ of size $U \times V$ is the data term, depending on input image $x, x\in \Re^{U \times V \times d}$ and  $U\times V \times d$ is the image width, height and depth respectively.
Both $\alpha(y_s)$ and $\beta(y_s)$ are weights associated with feature maps of dimension $U\times V$ extracted during the CNN training, same as $D(y_s)$. The terms associated with $\alpha$ and $\beta$ are first order and second order derivative of the polygon at $y_s$ defining length and curvature terms.   
$\sum_{u,v \in \Omega{(y)}}: \Omega{(y)} $ is the notation  to represent the pixels enclosed by the nodes of polygon $y$. 
$\sum_{u,v \in \Omega{(y)}}\kappa{(u,v)}$ is the summation of the pixel values of the kappa feature map enclosed within the polygon. This defines the kappa energy.  
We have modified the methodology used in \cite{marcos2018learning}. Their method uses custom CNN to learn Active Contour Model external and internal energy terms (refer to \autoref{eq1chap7}). They used the structured loss to train CNN (explained in \autoref{C7S3S5.1}). The complete details of their method, including active contour inference and experimental setup, could be found in their paper \cite{marcos2018learning}.
\subsection{Automatic Multiple Initialization}
\label{C7S3S4}
\begin{figure}[htbp]
\centering
\includegraphics[width=\textwidth ,height=2in]{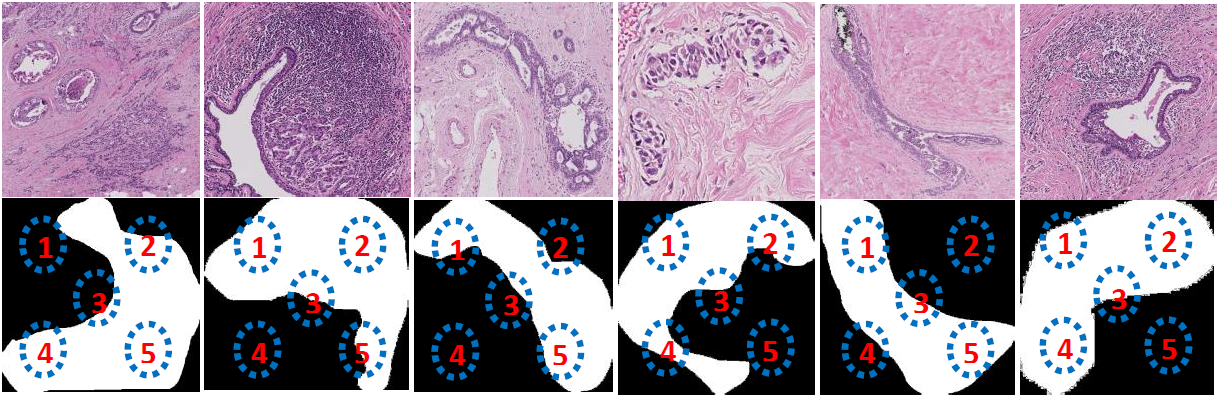}
\caption[Representation of modification in DSAC algorithm ]{Representation of modification in DSAC algorithm - Initial Polygon Selection Criterion: The first row comprise the labelled regions and the second row comprise their tumour mask. From left to right read as region label and the first polygon number with which we get the optimal final polygon, respectively: 1. Invasive, 5th, 2. Benign, 2nd, 3. In-situ, 5th, 4. Invasive, 1st, 5. Benign, 3rd, 6. Benign, 2nd or 4th}
\label{C7F2}
\end{figure}

We have modified the initialization of active contours by introducing multiple initial contours. There is a total of five contours initialized at four corners and one centre of the image. Each initial contour moves at each active contour inference iteration towards the minimum energy location. At the end of active contour iterations, the IoU over all the five predicted contours is calculated with the ground truth. The maximum IoU we obtain is then projected as the final predicted contour. This modification has been proposed to enhance the detection of objects that are slightly shifted towards the boundary of the image window rather than at the centre. This is explained through the illustration given in the \autoref{C7F2}

With multiple initializations of snake contour across the image window, there would be more chances of finding an optimal tumour boundary and minimize the convergence of snake at local maxima. This method is simple and robust for datasets with a single object per image. 
\subsection{U-Net training with structured SVM loss}
\label{C7S3S5}
In this method, energy terms are learned on the training sets instead of taking them as constants. Also, the structured loss used in the method is more suited for such complex datasets where both the target domain and the loss are more or less arbitrary. This means that the goal is not a simple target like a label or a number, but possibly a much more complicated object \cite{nowozin2011structured}. 
A non-trivial task of segmenting tumour boundaries is a suitable problem for structured prediction. Here the target mask differs significantly in local features such as size, shape, intensity, color and texture. If viewed as a pure segmentation problem, we could see that each possible snake iteration in the training set is provoked by varied values of internal and external forces (i.e., without a uniform pattern or range). This has several drawbacks when the loss is defined by a particular function like softmax, tanh, etc. Therefore, structured loss which considers the output (the segmentation mask) as a whole and not a set of arbitrary snake points is a more preferable choice. Moreover, structured prediction enables to conform relationship among multiple output variables (snake points) into a model (one output). 
\textbf{Terms}:
$X$ - Image  space\\
$Y$ - Output space\\
$y^i$ - positive ground truth polygon corresponding to $ith$ sample image $x^i$\\
$y$ - all the negative outputs where $y\neq y^i$ generated after every Active contour inference\\
$\hat{y}$ - predicted polygon \\
$\hat{y}^i$ - predicted polygon for $i^{th}$ sample \\
$\Delta (y,\hat{y})=\frac{y \cap \hat{y}}{y \cup \hat{y}}$ - task loss function or IoU \\
Energy $E(y)$ corresponding to output polygon $y$\\
Given a collection of ground truth pairs $(y^i, x^i) \in Y \times X, i=1,2,\ldots,N,$ and a task loss function $\Delta(y,\hat{y})$ where $y\in Y \wedge y\neq \hat{y}$, we would like to find CNN parameters $\omega$ such that by optimizing Eq. (1) and thus obtaining the inference result for $ith$ sample:
\bgroup                                  
\setlength{\abovedisplayskip}{0pt}
\begin{equation}
\hat{y}^i=arg\min_{y\in Y}E(y;\omega)
\end{equation}  
\egroup
one could expect a small $\Delta(y^i, \hat{y}^i)$. 
The problem becomes:
\bgroup                                  
\setlength{\abovedisplayskip}{0pt} 
\begin{equation}
\hat{\omega}= arg \min_{\omega} \sum_i \Delta(y^i, arg\min_{y\in Y}E(y;\omega))
\end{equation}
\egroup   

\subsubsection{Structured SVM loss}
\label{C7S3S5.1}
Since $\Delta(y^i, \hat{y}^i)$ could be a discontinuous function, this loss can be substituted by a continuous and convex function such as \textsc{hinge loss}.    
\begin{equation}
l(y^i;\omega)=max(0,\Delta(y^i,y)+E(y^i;\omega)-E(y;\omega)) 
\end{equation}
\begin{equation}
l(y^i;\omega)=max(0,\max_{y\in Y}(\Delta(y^i,y)+E(y^i;\omega)-E(y;\omega)))
\end{equation}
In Equation 7.5 the energy $E(y;\omega)$ corresponding to output $y$ decreases with every $omega$ update such that the difference between energy corresponding to the ground truth $E(y^i;\omega)$ and $E(y;\omega)$ is minimized. And, max over output space $Y$ is taken to maximize the margin between two energies such that when the Energy $E(y;\omega)$ decreases the task loss $\Delta(y^i,y)$ increases.\\
Now adding $l_2$ regularization and summing up for all training samples, hinge loss becomes the \textsc{max-margin formulation} which is our objective function:
\begin{equation}
L(Y;\omega)=\frac{1}{2}\parallel \omega \parallel^2 + C\sum_i \max(0,\max_{y\in Y}(\Delta(y^i,y)+E(y^i;\omega)-E(y;\omega)))
\end{equation}
where \textsc{Prediction function}  is defined as:
\begin{equation}
\hat{y}^i=arg\max_{y\in Y}(\Delta(y^i,y)+E(y^i;\omega)-E(y;\omega))
\end{equation}
If the constant (energy corresponding to ground truth $E(y^i;\omega)$ is dropped from the above equation, we obtain
\begin{equation}
\hat{y}^i=arg\max_{y\in Y}(\Delta(y^i,y)-E(y;\omega))  
\end{equation}
Once we obtain $\hat{y}^i$   then the Objective with stochastic approx for randomly chosen data point $i$ becomes:  
\begin{equation}
L(Y;\omega)=\frac{1}{2}\parallel \omega \parallel^2 + C\max(0,\Delta(y^i,\hat{y}^i)+E(y^i;\omega)-E(\hat{y}^i;\omega)))
\end{equation}         
Since $L(Y;\omega)$ is not differentiable, gradients cannot be calculated. hence, we compute subgradient as:
\begin{equation}
\frac{\partial L(Y;\omega)}{\partial \omega}= \omega + C \frac{\partial}{\partial \omega} \max(0,\Delta(y^i,\hat{y}^i)+E(y^i;\omega)-E(\hat{y}^i;\omega)))
\end{equation}
where; 
\begin{gather}
\frac{\partial}{\partial \omega} (\max(0,\Delta(y^i,\hat{y}^i)-E(\hat{y}^i;\omega)+E(y^i;\omega))) =
\left\{
    \begin{array}{ll}
        \frac{\partial E(y^i;\omega)}{\partial \omega} - \frac{\partial E(\hat{y}^i;\omega)}{\partial \omega}  & \mbox{; if } E(y^i;\omega)- E(\hat{y}^i;\omega)  < \Delta(y^i,\hat{y}^i) \\
        0 & \mbox{; if } E(y^i;\omega)- E(\hat{y}^i;\omega)  = \Delta(y^i,\hat{y}^i) \\
        0 & \mbox{; if } E(y^i;\omega)- E(\hat{y}^i;\omega)  > \Delta(y^i,\hat{y}^i)
    \end{array}
\right.
\end{gather}
So, following is the subgradient with respect to $\omega$:
\begin{equation}
\left\{
    \begin{array}{ll}
        \mbox{if  } E(y^i;\omega)- E(\hat{y}^i;\omega) < \Delta(y^i,\hat{y}^i)  & \mbox{; } \omega + C(\frac{\partial E(y^i;\omega)}{\partial \omega} - \frac{\partial E(\hat{y}^i;\omega)}{\partial \omega}) \\
        \mbox{else  } & \mbox{; } \omega+0
    \end{array}
\right.
\end{equation} 
The algorithm \ref{alg0chap7_2} describes the our methods training for U-Net architecture. 
\setlength{\textfloatsep}{1pt}
\begin{algorithm}
\caption{U-Net training}
\begin{algorithmic}
\State $X$ - Image space \par
\State $Y$ - Output space\par
\State $y^i$ - positive ground truth polygon corresponding to $ith$ sample image $x^i$\par
\State $y$ - all the negative outputs where $y\neq y^i$ generated after every Active contour inference\par
\State $N$ - total number of samples in a batch. 
\State $\omega$ - U-Net parameter to be updated
\State $\hat{y}$ - predicted polygon \par
\State $\hat{y}^i$ - predicted polygon for $i^{th}$ sample \par
\State $\Delta (y,\hat{y})=\frac{y \cap \hat{y}}{y \cup \hat{y}}$ - task loss function or IoU \par
\State input number of iterations T, step size $\eta$ for $t=1,\ldots,T$ \par
\State Energy $E(y;\omega)$ corresponding to output polygon $y$  \par
\State Energy $E(\hat{y}^i;\omega)$ corresponding to predicted polygon $\hat{y}^i$ for $ith$ sample \par 
\State regularizer $C$ \par
\State \textbf{Initialize} $\omega\leftarrow \vec{0}$
\For{$t=1,\ldots,T$}\par
    \For {$i=1,\ldots,N$} \par 
      \State $\hat{y}^{i} \gets arg\max_{y\in Y}(\Delta(y^i,y)-E(y;\omega))$ \par
      \If {$E(y^i;\omega)- E(\hat{y}^i;\omega)  < \Delta(y^i,\hat{y}^i)$}
      \State $v^i \gets \frac{\partial E(y^i;\omega)}{\partial \omega} - \frac{\partial E(\hat{y}^i;\omega)}{\partial \omega}$ \par
      \Else 
      \State $v^i \gets 0$ \par
      \EndIf
    \EndFor \par
\State $\omega \gets \omega - \eta (\omega + \frac{C}{N}\sum_i^N v^i)$ \par    
\EndFor
\end{algorithmic}
\label{alg0chap7_2}
\end{algorithm}
\section{Experimental Setup and Results}
\label{C7S4}
For segmentation task, the dataset comprises the labelled regions from the WSIs which were resized to ($512 \times 512$) for segmenting the suspected tumour. In this case, normal patches were not included in the segmentation dataset.

\subsection{Dataset Preparation and Usage}
\label{C7S4S1}
\begin{figure}[H]
\centering
\includegraphics[width=\textwidth]{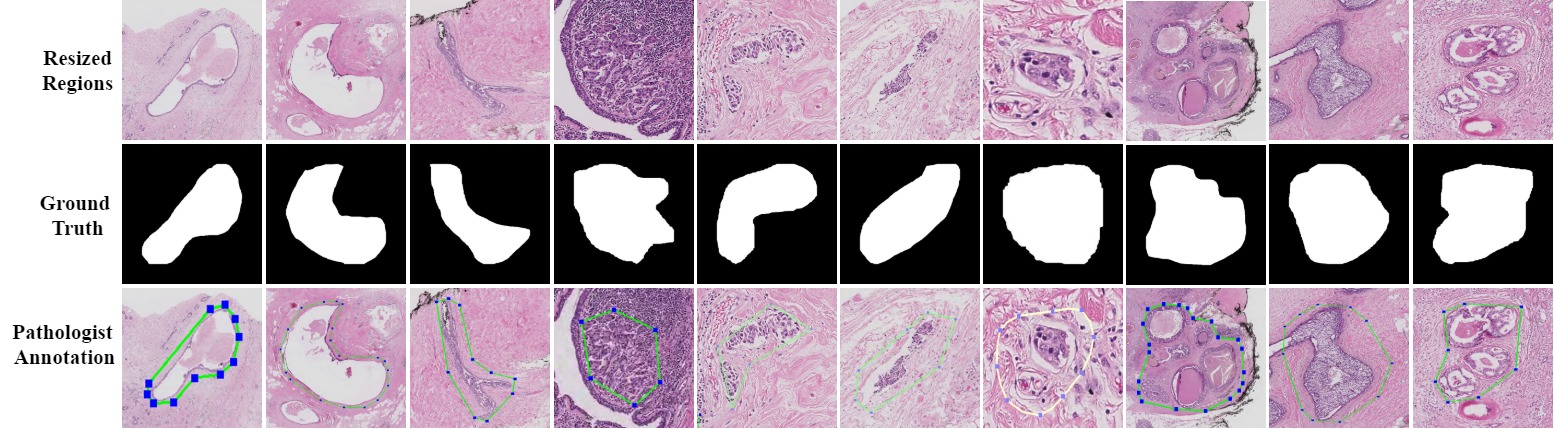}
\caption[WSI region samples with their corresponding ground truth mask]{WSI region samples with their corresponding ground truth mask generated through polygonal annotation by the pathologist \cite{bach2018}.}
\label{C7F3}
\end{figure}

The ICIAR BACH 2018 challenge published a breast WSI tumour dataset \cite{bach2018}. The dataset contains ten annotated WSIs for training. They did not however revealed the test annotation. Therefore, we have worked only on the tumour regions extracted from the ten WSIs. The WSI contains three annotated classes- Benign, Invasive, \textit{In situ}. Using the annotation coordinates in ground truth files, we calculated the bounding box dimensions around each annotated tumour region. We then increased the bounding box dimensions of the annotated regions by 40\% to increase the background area around the tumour mask. From the ten WSIs, total 56 Benign, 100 Invasive, and 60 \textit{In situ} regions were extracted. Each extracted region were of arbitrary dimensions ranging between 20,000 pixels to 196 pixels across height and width. To reduce the computational complexity and make the images of even dimensions, all the 216 regions were resized to $512 \times 512$ with 3 RGB color channels. \autoref{C7F3} shows the dataset samples with ground truth and corresponding pathologist annotation in each row, respectively. The polygonal annotation is expected to be achieved through active contour inference that moves towards the edge of the tumour while training on ground truth mask. From the figure, we could see how arbitrary and heterogeneous shapes are roughly annotated by the pathologist for only detection purpose. 

\subsection{Experiments}
\label{C7S4S2}
We did several experiments for preliminary analysis of our adopted method for breast tumour segmentation. The main method has been compared with the original DSAC method proposed in  \cite{marcos2018learning}, FCN16, and SegNet. Some ablation studies have been done to show the variation in results with hyperparameter changes such as the number of layers, optimizer, and learning rate. In the main method, we used Adam Optimizer with a learning rate of $10^4$ and five layers in an encoder and four layers in decoder branch.  We evaluated the performance of the method on IoU averaged over all test images. The dataset is randomly divided into 150 training images and 66 test images. 
\begin{table}[H]
\renewcommand{\arraystretch}{1.5}
\caption[Comparative performance evaluation over average IoU on the BACH dataset]{Comparative performance evaluation over average IoU on the test set of ICIAR BACH 2018 dataset}
\label{C7T1}
\centering
\begin{tabular}{|c|c|c|}
\hline
\textbf{Model}&\textbf{Method}&\textbf{Average IoU(\%)}\\
\hline
\multirow{5}{*}{\textbf{Semantic}}&\textbf{FCN16}&51.72\\
\cline{2-3}
&\textbf{SegNet}&71.65\\
\cline{2-3}
&\textbf{UNet}&44.69\\
\cline{2-3}
&\textbf{UNet+DenseNet}&51.08\\
\cline{2-3}
&\textbf{U-Net+Resnet}&77.13\\
\hline
\multirow{4}{*}{\textbf{\shortstack{Hybrid models}}}&\textbf{original DSAC}&56.09\\
\cline{2-3}
&\textbf{\shortstack{original DSAC\\ with multiple initializations}}&58.62\\
\cline{2-3}
&\textbf{\shortstack{original DSAC\\ with multiple initializations\\ and ResBlock}}&60.07\\
\cline{2-3}
&\textbf{U-Net-ResNet-ACM (ours)}&76.45\\
\hline
\end{tabular}
\end{table}

\begin{sidewaysfigure}[htbp]
\centering
\includegraphics[width=\textwidth]{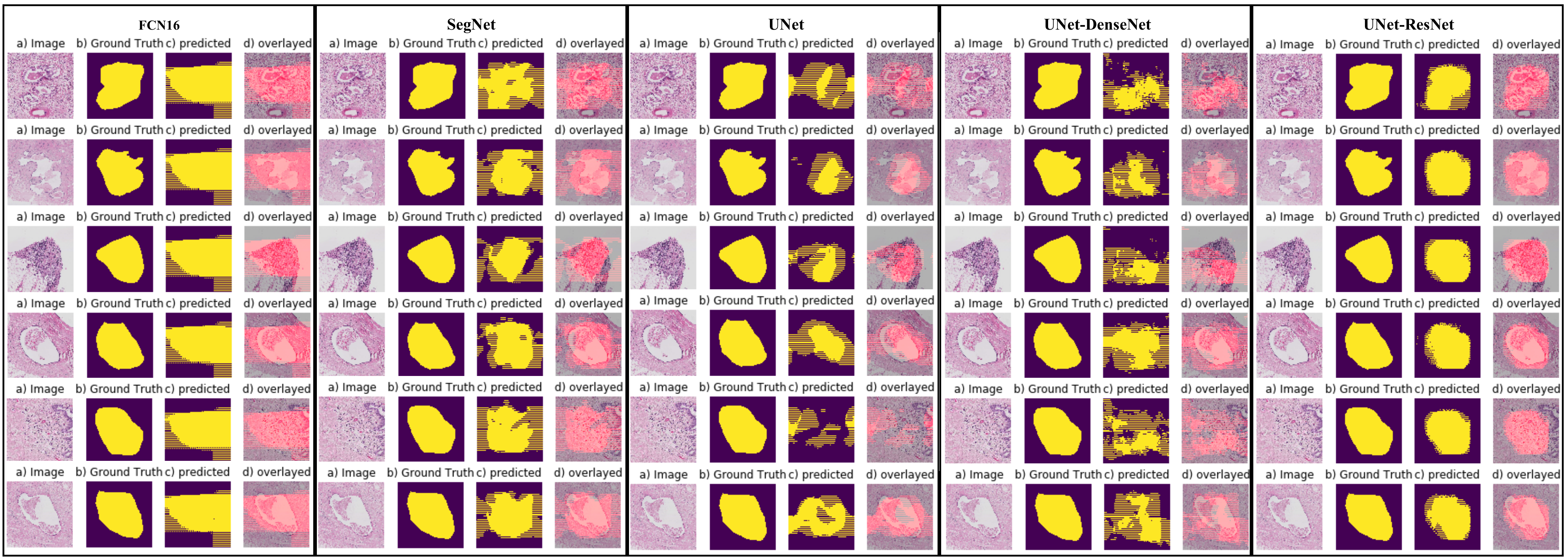}
\caption[Comparative semantic Segmentation output of different models]{Comparative semantic Segmentation output of different models. The first column in each model shows the original image, the second column shows the ground truth, the third column  illustrates the segmented output, and the fourth column shows the overlayed output on original image}
\label{C7F4}
\end{sidewaysfigure}

\begin{sidewaysfigure}[htbp]
\centering
\includegraphics[width=\textwidth]{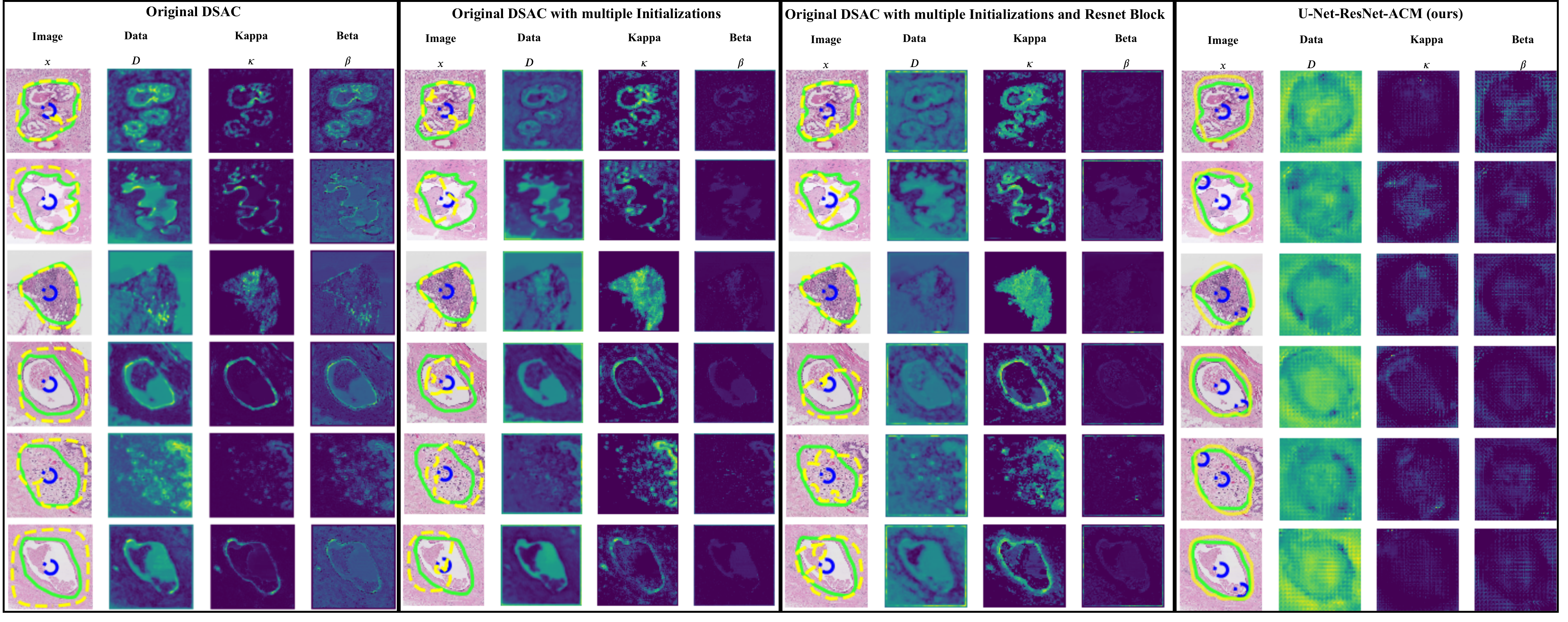}
\caption[Comparative hybrid Segmentation output of different models]{Comparative hybrid Segmentation output of different models. In the first column of each model, the initial contours are in blue and the obtained result in yellow, with the ground truth in green colour. The second column shows the Data term $D(x)$, where the boundary can be seen as a region with lower energy. The third column is the balloon terms $\kappa$ which highlights the region within the boundary, and the fourth column shows the $\beta$ feature map where we notice the curvature areas in the original image are penalized.}
\label{C7F5}
\end{sidewaysfigure}

The dataset was tested with original DSAC CNN backbone with and without multiple initializations. The average IoU obtained validated that with multiple initializations, the detection performance of active contour has increased. We then introduced ResNet blocks in the original CNN backbone to test whether there is an improvement with ResNet identity blocks. We observed that the IoU has increased from 59.98\% to 61.32\% with ResNet blocks in the original CNN backbone architecture followed in original DSAC. The observed results strengthen the choice of including ResNet blocks and multiple initializations in our framework. Further, we replaced the CNN model with U-Net and ResNet blocks with multiple initializations as our final proposed model. The proposed model is tested with semantic segmentation networks like SegNet, FCN16, original U-Net. The original U-Net was then further enhanced with ResNet and DenseNet blocks, respectively. The \autoref{C7T1} shows a comparison between different models with our proposed model. From the observed IoU, we could deduce that the choice of deep learning backbone affects the final performance of the active contour inference over the image. When we compared the results of semantic segmentation models with Active Contour enhanced hybrid semantic models, except the U-Net + ResNet model, we observed incremental improvement with hybrid approaches. The results hence strengthen the idea of annotating medical datasets with such models to imitate pathologist like annotations instead of using semantic segmentation models which are not usually useful in clinical settings. \autoref{C7F4} shows the results obtained after semantic segmentation of dataset test images using state-of-the-art semantic models. 
Further, the results obtained from hybrid segmentation models is illustrated in \autoref{C7F5}.
\section{Discussion}
\label{C7S5}
Through this work, we aimed to apply the deep learning trained active contour segmentation on complex histopathology breast tumour dataset. The dataset is roughly annotated by the pathologists for marking tumour regions. The dataset is not explicitly annotated for segmentation purpose. This makes the tasks more challenging. Hence, The base model is enhanced by introducing ResNet blocks for recovering information loss during the downsampling and upsampling operations in the network. The final segmentation results, as shown, prove the sensitivity of the network with ResNet blocks.  For marking the boundary of the tumour with a polygon just like the pathologist does, the active contour algorithm with locally learned priors is added to the segmentation model. The active contour is moved to detect the boundaries of the tumour using the strong local priors learned by the U-Net-ResNet deep learning model. Multiple automatic initializations proved to be the critical factor to improve the detection performance of the algorithm.
We started experiments with the new state of the art semantic and instance segmentation algorithms. The results in \autoref{C7T1} show the comparison of semantic segmentation methods like UNeT, FCN, and SegNet with our modified model application on histopathology dataset. Clearly, we could see that pixel-wise classification with pixels as segmenting unit did not work well with our dataset and our intuition about heterogeneity causing misclassification of such a small unit like a pixel is correct. The problem of heterogeneity within objects, their shape, size, coarseness differ so widely in histopathology images that segmentation of coarsely annotated regions of interest like tumours causes either over-segmentation or under segmentation of pixels. For example, in the case of Ductal Carcinoma In Situ (DCIS) of the breast, there are at least five subtypes namely, DCIS: micropapillary, DCIS: Cribriform, DCIS: Cribriform with microcalcifications, and DCIS: Apocrine and DCIS: Comedonecrosis. All these subtypes have different shapes, distribution of nuclei and the presence of substructures. So, without subtype annotations, a computer algorithm treats every subtype as a different class. If the matching instance is not present within that slide, the algorithm fails to recognize the instance of DCIS tumour.  Therefore, without extensive manual annotation of each substructure, semantic and instance segmentation models largely fail in such scenarios. Our method, with active contour inference, increase the object awareness through active feature learning and contour displacement within an end-to-end network. Thus, it made the overall network more sensitive to object features. And with it, we achieved our aim of polygonal annotations just as the pathologist would mark for cancer detection in histopathological images.
\section{Conclusion}
\label{C7S6}
We have applied the state-of-the-art deep structured active contour model on a medical dataset to imitate pathologist annotations for tumours in the breast histopathology dataset. For the task, we introduced semantic segmentation model U-Net enhanced by ResNet blocks to learn local information priors for active contour inference actively. The initial active contours acted as object identifiers which helped to improve the network performance for heterogeneous data. The future work would be to enhance the segmentation performance at WSI level so that the pathologist like annotations could be done for both medical and educational purposes. 
\section*{Acknowledgement}
This research was carried out in the Indian Institute of Information Technology, Allahabad and supported by the Ministry of Human Resource and Development, Government of India. We would also like to acknowledge the support and guidance of Dr Hwee Kuan Lee, Principal Investigator, A*STAR Bioinformatics Institute, Singapore.  We are also grateful to the NVIDIA corporation for supporting our research in this area by granting us TitanX (PASCAL) GPU. 

\section*{Declaration of Interest}
Declarations of interest: none

\bibliographystyle{elsarticle-num}
\bibliography{tumourseg}

\begin{thebibliography}{10}
\expandafter\ifx\csname url\endcsname\relax
  \def\url#1{\texttt{#1}}\fi
\expandafter\ifx\csname urlprefix\endcsname\relax\def\urlprefix{URL }\fi
\expandafter\ifx\csname href\endcsname\relax
  \def\href#1#2{#2} \def\path#1{#1}\fi

\bibitem{ronneberger2015u}
O.~Ronneberger, P.~Fischer, T.~Brox, U-net: Convolutional networks for
  biomedical image segmentation, in: International Conference on Medical image
  computing and computer-assisted intervention, Springer, 2015, pp. 234--241.

\bibitem{li2017multi}
J.~Li, K.~V. Sarma, K.~C. Ho, A.~Gertych, B.~S. Knudsen, C.~W. Arnold, A
  multi-scale u-net for semantic segmentation of histological images from
  radical prostatectomies, in: AMIA Annual Symposium Proceedings, Vol. 2017,
  American Medical Informatics Association, 2017, p. 1140.

\bibitem{qu2019joint}
H.~Qu, G.~Riedlinger, P.~Wu, Q.~Huang, J.~Yi, S.~De, D.~Metaxas, Joint
  segmentation and fine-grained classification of nuclei in histopathology
  images, in: 2019 IEEE 16th International Symposium on Biomedical Imaging
  (ISBI 2019), IEEE, 2019, pp. 900--904.

\bibitem{alom2018nuclei}
M.~Z. Alom, C.~Yakopcic, T.~M. Taha, V.~K. Asari, Nuclei segmentation with
  recurrent residual convolutional neural networks based u-net (r2u-net), in:
  NAECON 2018-IEEE National Aerospace and Electronics Conference, IEEE, 2018,
  pp. 228--233.

\bibitem{sirinukunwattana2017gland}
K.~Sirinukunwattana, J.~P. Pluim, H.~Chen, X.~Qi, P.-A. Heng, Y.~B. Guo, L.~Y.
  Wang, B.~J. Matuszewski, E.~Bruni, U.~Sanchez, et~al., Gland segmentation in
  colon histology images: The glas challenge contest, Medical image analysis 35
  (2017) 489--502.

\bibitem{graham2019mild}
S.~Graham, H.~Chen, J.~Gamper, Q.~Dou, P.-A. Heng, D.~Snead, Y.~W. Tsang,
  N.~Rajpoot, Mild-net: minimal information loss dilated network for gland
  instance segmentation in colon histology images, Medical image analysis 52
  (2019) 199--211.

\bibitem{zeng2019ric}
Z.~Zeng, W.~Xie, Y.~Zhang, Y.~Lu, Ric-unet: An improved neural network based on
  unet for nuclei segmentation in histology images, Ieee Access 7 (2019)
  21420--21428.

\bibitem{bentaieb2016topology}
A.~BenTaieb, G.~Hamarneh, Topology aware fully convolutional networks for
  histology gland segmentation, in: International Conference on Medical Image
  Computing and Computer-Assisted Intervention, Springer, 2016, pp. 460--468.

\bibitem{cui2018deep}
Y.~Cui, G.~Zhang, Z.~Liu, Z.~Xiong, J.~Hu, A deep learning algorithm for
  one-step contour aware nuclei segmentation of histopathological images, arXiv
  preprint arXiv:1803.02786 (2018).

\bibitem{oda2018besnet}
H.~Oda, H.~R. Roth, K.~Chiba, J.~Sokoli{\'c}, T.~Kitasaka, M.~Oda, A.~Hinoki,
  H.~Uchida, J.~A. Schnabel, K.~Mori, Besnet: boundary-enhanced segmentation of
  cells in histopathological images, in: International Conference on Medical
  Image Computing and Computer-Assisted Intervention, Springer, 2018, pp.
  228--236.

\bibitem{chan2001active}
T.~F. Chan, L.~A. Vese, Active contours without edges, IEEE Transactions on
  image processing 10~(2) (2001) 266--277.

\bibitem{he2016deep}
K.~He, X.~Zhang, S.~Ren, J.~Sun, Deep residual learning for image recognition,
  in: Proceedings of the IEEE conference on computer vision and pattern
  recognition, 2016, pp. 770--778.

\bibitem{huang2017densely}
G.~Huang, Z.~Liu, L.~Van Der~Maaten, K.~Q. Weinberger, Densely connected
  convolutional networks, in: Proceedings of the IEEE conference on computer
  vision and pattern recognition, 2017, pp. 4700--4708.

\bibitem{marcos2018learning}
D.~Marcos, D.~Tuia, B.~Kellenberger, L.~Zhang, M.~Bai, R.~Liao, R.~Urtasun,
  Learning deep structured active contours end-to-end, in: Proceedings of the
  IEEE Conference on Computer Vision and Pattern Recognition, 2018, pp.
  8877--8885.

\bibitem{mehta2018net}
S.~Mehta, E.~Mercan, J.~Bartlett, D.~Weaver, J.~G. Elmore, L.~Shapiro, Y-net:
  joint segmentation and classification for diagnosis of breast biopsy images,
  in: International Conference on Medical Image Computing and Computer-Assisted
  Intervention, Springer, 2018, pp. 893--901.

\bibitem{badrinarayanan2017segnet}
V.~Badrinarayanan, A.~Kendall, R.~Cipolla, Segnet: A deep convolutional
  encoder-decoder architecture for image segmentation, IEEE transactions on
  pattern analysis and machine intelligence 39~(12) (2017) 2481--2495.

\bibitem{long2015fully}
J.~Long, E.~Shelhamer, T.~Darrell, Fully convolutional networks for semantic
  segmentation, in: Proceedings of the IEEE conference on computer vision and
  pattern recognition, 2015, pp. 3431--3440.

\bibitem{ackerman1971pathology}
L.~V. Ackerman, J.~Rosai, The pathology of tumors, part one: introduction,
  precancerous lesions, benign lesions that resemble cancer, CA: A Cancer
  Journal for Clinicians 21~(3) (1971) 162--173.

\bibitem{he2017mask}
K.~He, G.~Gkioxari, P.~Doll{\'a}r, R.~Girshick, Mask r-cnn, in: Proceedings of
  the IEEE international conference on computer vision, 2017, pp. 2961--2969.

\bibitem{girshick2015fast}
R.~Girshick, Fast r-cnn, in: Proceedings of the IEEE international conference
  on computer vision, 2015, pp. 1440--1448.

\bibitem{ren2015faster}
S.~Ren, K.~He, R.~Girshick, J.~Sun, Faster r-cnn: Towards real-time object
  detection with region proposal networks, in: Advances in neural information
  processing systems, 2015, pp. 91--99.

\bibitem{mehta2018learning}
S.~Mehta, E.~Mercan, J.~Bartlett, D.~Weaver, J.~Elmore, L.~Shapiro, Learning to
  segment breast biopsy whole slide images, in: 2018 IEEE Winter Conference on
  Applications of Computer Vision (WACV), IEEE, 2018, pp. 663--672.

\bibitem{geccer2016detection}
B.~Ge{\c{c}}er, Detection and classification of breast cancer in whole slide
  histopathology images using deep convolutional networks, Diss. Bilkent
  University 1 (2016).

\bibitem{kainz2015semantic}
P.~Kainz, M.~Pfeiffer, M.~Urschler, Semantic segmentation of colon glands with
  deep convolutional neural networks and total variation segmentation, arXiv
  preprint arXiv:1511.06919 (2015).

\bibitem{chen2016dcan}
H.~Chen, X.~Qi, L.~Yu, P.-A. Heng, Dcan: deep contour-aware networks for
  accurate gland segmentation, in: Proceedings of the IEEE conference on
  Computer Vision and Pattern Recognition, 2016, pp. 2487--2496.

\bibitem{xu2010weighted}
J.~Xu, A.~Janowczyk, S.~Chandran, A.~Madabhushi, A weighted mean shift,
  normalized cuts initialized color gradient based geodesic active contour
  model: applications to histopathology image segmentation, in: Medical Imaging
  2010: Image Processing, Vol. 7623, International Society for Optics and
  Photonics, 2010, p. 76230Y.

\bibitem{xu2011high}
J.~Xu, A.~Janowczyk, S.~Chandran, A.~Madabhushi, A high-throughput active
  contour scheme for segmentation of histopathological imagery, Medical image
  analysis 15~(6) (2011) 851--862.

\bibitem{ali2012integrated}
S.~Ali, A.~Madabhushi, An integrated region-, boundary-, shape-based active
  contour for multiple object overlap resolution in histological imagery, IEEE
  transactions on medical imaging 31~(7) (2012) 1448--1460.

\bibitem{xu2019convolutional}
J.~Xu, L.~Gong, G.~Wang, C.~Lu, H.~Gilmore, S.~Zhang, A.~Madabhushi,
  Convolutional neural network initialized active contour model with adaptive
  ellipse fitting for nuclear segmentation on breast histopathological images,
  Journal of Medical Imaging 6~(1) (2019) 017501.

\bibitem{khvostikov2019trainable}
A.~Khvostikov, A.~Krylov, I.~Mikhailov, P.~Malkov, Trainable active contour
  model for histological image segmentation, Scientific Visualization 11~(3)
  (2019).

\bibitem{nowozin2011structured}
S.~Nowozin, C.~H. Lampert, et~al., Structured learning and prediction in
  computer vision, Foundations and Trends{\textregistered} in Computer Graphics
  and Vision 6~(3--4) (2011) 185--365.

\bibitem{bach2018}
[dataset] iciar breast cancer histology images (bach) 2018,
  \url{https://iciar2018-challenge.grand-challenge.org/Home/} (November 2018).

\end{thebibliography}

\end{document}